
\hsize=14 cm \vsize=20.8 cm \tolerance=400
\hoffset=2cm
\font\trm = cmr10 scaled \magstep3
\font\srm = cmr10 scaled \magstep2

\voffset=2cm
\scriptscriptfont0 =\scriptfont0
\scriptscriptfont1 =\scriptfont1

\def\d{\partial}

\def\sqr#1#2{{\vcenter{\vbox{\hrule height.#2pt\hbox{\vrule width.#2pt
height#1pt \kern#1pt \vrule width.#2pt}\hrule height.#2pt}}}}
\def\w{\mathchoice\sqr45\sqr45\sqr{2.1}3\sqr{1.5}3\,}
\def\fii{\varphi}

\def\psq{{\overline{\psi}}}

\def\=d{\,{\buildrel\rm def\over =}\,}

\def\i3p{\p32\int d^3p}

\def\ds{\hbox{\rlap/$\partial$}}

\def\As{A\hbox to 1pt{\hss /}}
\def\np4{\int d^4p_1\cdots d^4p_{n-1}\, }

\def\supp{{\rm supp}\, }

\def\nx4{\int d^4x_1\ldots d^4x_n\, }
\def\xnn{x_1,\ldots ,x_n}

\def\kon#1#2{\vbox{\halign{##&&##\cr
\lower4pt\hbox{$\scriptscriptstyle\vert$}\hrulefill &
\hrulefill\lower4pt\hbox{$\scriptscriptstyle\vert$}\cr $#1$&
$#2$\cr}}}

\def\konv#1#2#3{\hbox{\vrule height12pt depth-1pt}
\vbox{\hrule height12pt width#1cm depth-11.6pt}
\hbox{\vrule height6.5pt depth-0.5pt}
\vbox{\hrule height11pt width#2cm depth-10.6pt\kern5pt
      \hrule height6.5pt width#2cm depth-6.1pt}
\hbox{\vrule height12pt depth-1pt}
\vbox{\hrule height6.5pt width#3cm depth-6.1pt}
\hbox{\vrule height6.5pt depth-0.5pt}}
\def\konu#1#2#3{\hbox{\vrule height12pt depth-1pt}
\vbox{\hrule height1pt width#1cm depth-0.6pt}
\hbox{\vrule height12pt depth-6.5pt}
\vbox{\hrule height6pt width#2cm depth-5.6pt\kern5pt
      \hrule height1pt width#2cm depth-0.6pt}
\hbox{\vrule height12pt depth-6.5pt}
\vbox{\hrule height1pt width#3cm depth-0.6pt}
\hbox{\vrule height12pt depth-1pt}}

\def\konw#1#2#3{\hbox{\vrule height12pt depth-1pt}
\vbox{\hrule height12pt width#1cm depth-11.6pt}
\hbox{\vrule height6.5pt depth-0.5pt}
\vbox{\hrule height12pt width#2cm depth-11.6pt \kern5pt
      \hrule height6.5pt width#2cm depth-6.1pt}
\hbox{\vrule height6.5pt depth-0.5pt}
\vbox{\hrule height12pt width#3cm depth-11.6pt}
\hbox{\vrule height12pt depth-1pt}}

\def\i{{\rm int}}
\def\su{\sum_{n=1}^\infty}

\def\snn{\su {1\over n!}\nx4}
\def\r{{\rm ret}}
\def\a{{\rm av}}
\def\ra{{{\rm\scriptstyle ret}\atop{\rm\scriptstyle av}}}

\def\m3{{\mu_1\mu_2\mu_3}}

\def\p{{(+)}}

\nopagenumbers
\vbox to 4cm{ }
\centerline{\trm Gauge invariance of massless QED} \vskip 2cm
\centerline{\srm M.D\"utsch, T.Hurth and G.Scharf} \vskip 0.5cm
\centerline{\it
Institut f\"ur Theoretische Physik der Universit\"at Z\"urich}
\centerline{\it Winterthurerstr. 190, CH-8057 Z\"urich, Switzerland}\vskip 3cm
A simple general proof of gauge invariance in QED is given in the
framework of causal perturbation theory. It illustrates a method which
can also be used in non-abelian gauge theories.
\vfill\eject
\pageno=1
\headline={\tenrm\ifodd\pageno\hss\folio\else\folio\hss\fi}
\footline={\hss}
 The existing rigorous proofs of gauge invariance in perturbative QED
[1, 2] are rather involved due to ultraviolet and infrared problems. The
latter are severe enough, so that these proofs do not apply to the case
of massless fermions. We present a simple general proof to fill this
gap. It covers both the massless and massive case.

The general idea of proving gauge invariance is well-known: Gauge
invariance can only be spoiled by certain local anomalous terms, one
then tries to absorb those terms by (finite!) renormalizations. It is
non-trivial  that this is possible, because there are more anomalies
than there is freedom in normalization. Some further property of QED
must be used. Here different possibilities exist. We will show that
charge conjugation (C-invariance) is sufficient for our purpose.
We do not discuss the problem of fixing {\it all} normalization
constants of the theory by further physical conditions. This would
require a careful study of the infrared problems in the adiabatic limit,
which lies beyond the scope of this note. Such problems were studied by
Hurd [6] using the methods of [1]. In our approach to gauge invariance,
the infrared problem plays no role.

We work with causal perturbation theory [3-5] where there is no
ultraviolet problem and only well defined, finite quantities appear.
In this
framework, gauge invariance is expressed as follows [3]: Let $g(x)\in S({\bf
R^4})$ be a test function (switching function) and $S(g)$ the
perturbatively defined S-matrix
$$S(g)=1+\snn T_n(\xnn)g(x_1)\ldots g(x_n),\eqno(1)$$
and let the $n$-point function $T_n$ be normally ordered with respect to
the (free) photon operators
$$T_n(\xnn)=\sum_{l=0}^n\>\sum_{1\le k_1<\ldots <k_l\le n}\>
t_{k_1\ldots k_l}^{\mu_1\ldots\mu_l}(\xnn)$$
$$\times\, :\,A_{\mu_1}(x_{k_1})\ldots A_{\mu_l}(x_{k_l})\, :.\eqno(2)$$
Then we must have
$${\d\over\d x_{k_j}^{\mu_j}}\, t_{k_1\ldots k_l}^{\mu_1\ldots\mu_l}
(\xnn)=0,\eqno(3)$$
for all $1\le l\le n$, all $1\le j\le l$, all $1\le k_1<\ldots <k_l\le n$
and all $(\xnn)\in {\bf R^{4n}}$. The $t$'s in (2) contain the Fermi
operators: $t_{k_1\ldots k_l}^{\mu_1\ldots\mu_l}$ is the sum of all
graphs of order $n$ with external photon lines at the vertices $x_{k_1},
\ldots x_{k_l}$ and no other external photon lines, the external
fermions being arbitrary. It follows from (2) that $t_{k_1\ldots k_l}
^{\mu_1\ldots\mu_l}$ is symmetrical in $(x_{k_1},\mu_1)\ldots,
(x_{k_l},\mu_l)$.

It is our aim to prove gauge invariance (3) by induction on $n$.
For the beginning of the induction we refer to [3], Chap.3.11, and we
consider $n>3$ here.
Let us assume that (3) holds for all $m$-point distributions with
$m\le n-1$. Going from $n-1$ to $n$ according to the inductive
construction [3], we have first to form
$$R'_n(\xnn)=\sum_XT_{n-n_1}(Y,x_n)\tilde T_{n_1}(X),\eqno(4)$$
where $\tilde T_{n_1}$ comes from the perturbation expansion (1) of
$S(g)^{-1}$. Each term in (4) is a product of $T_m$'s with $m\le n-1$
and disjoint arguments. In virtue of the induction assumption, each term
is gauge invariant, because the normal ordering in the photon operators
does not affect it. The same is true for
$$A'_n(\xnn)=\sum_X\tilde T_{n_1}(X)T_{n-n_1}(Y,x_n),\eqno(5)$$
and for
$$D_n=R'_n-A'_n.\eqno(6)$$
This distribution has causal support with respect to $x_n$:
$$\supp D_n\subseteq\Gamma_+^n(x_n)\cup\Gamma_-^n(x_n),\eqno(7)$$
$$\Gamma_\pm^n(x_n)=\{(\xnn)\>|\>x_j\in V^\pm(x_n),\,\forall j=1,\ldots
n-1\},\eqno(8)$$
where $V^\pm (x)$ is the closed forward or backward cone of $x$,
respectively. The essential step in the inductive construction is the
splitting of $D_n=R_n-A_n$ into a retarded and advanced part.
There remains to prove that gauge invariance is preserved under
this operation. The final steps $T_n=R_n-R'_n$ and symmetrization with
respect to $x_n$ do not affect gauge invariance.

Our starting point is the relation
$$\d_\nu d^\nu\=d{\d\over\d x_n^\nu}d^{\mu_1\ldots\mu_{l-1}\nu}_{k_1
\ldots k_l}(x_1,\ldots x_{n-1},x_n)=0,\eqno(9)$$
where we have introduced a shorthand notation and have taken
$x_{k_j}=x_n, \mu_j=\nu$ for simplicity. Since the retarded part is
equal to
$$r^\nu(\xnn)=\cases{d^\nu(\xnn)&on $\quad\Gamma^n_+(x_n)\setminus
(x_n,\ldots x_n)$\cr 0& on $\quad\Gamma^n_+(x_n)^C,$\cr}\eqno(10)$$
we conclude from (9) that $\d_\nu r^\nu$ can only have a point support:
$$\supp\d_\nu r^\nu(\xnn)\subseteq (x_n,\ldots x_n).\eqno(11)$$
We decompose
$$d^\nu =\sum_{f=0}^n d_f^\nu\quad,\quad r^\nu=\sum_{f=0}^n r^\nu_f
\eqno(12)$$
into contributions of all graphs of order $n$ with $2\cdot f$ external
fermion lines and $l$ external photon lines. According to (9), we
must separately have
$$\d_\nu d^\nu_{f'}(\xnn)=0\quad,\quad\forall 0\le f'\le n.\eqno(13)$$

It is important to note that there are no terms with derivative $\d_\nu$
acting on a
Fermi field operator for the following reason: This derivative comes
from an external photon operator $A^\nu(x_n)$. If there is also a Fermi
operator $\psq(x_n)$, the vertex $x_n$ is connected with the rest of the
graph by a single fermion line, represented by a Fermi propagation function
$S_\r(x_n-x_j)$, $S_\a$ or $S_F$. Such a contribution can be reduced by means
of the Dirac equation
$${\d\over\d x_n^\nu}\Bigl(\psq(x_n)\gamma^\nu S_\ra (x_n-x_j)\Bigl)
=i\psq(x_n)\delta^4(x_n-x_j),\eqno(14)$$
$${\d\over\d x_n^\nu}\Bigl(S_\ra (x_j-x_n)\gamma^\nu\psi (x_n)\Bigl)
=-i\delta^4(x_j-x_n)\psi(x_n).\eqno(15)$$

Next we carry out the normal product decomposition and split all
numerical distributions.
Now (11) must also hold for every $r^\nu_{f'}$  separately. Using a
well-known theorem on distributions with point-like support, we arrive
at
$$\d_\nu r^\nu_{f'}(\xnn)=\sum_g:\,\psq(x_{i_1})\ldots\psq(x_{i_{f'}})
\Bigl(\,\sum_{|a|\le\omega(g)+1} K^g_aD^a\delta (x_1-x_n)\ldots$$
$$\ldots\delta(x_{n-1}-x_n)\Bigl)\psi(x_{j_1})\ldots\psi(x_{j_{f'}})
\,:\>,\eqno(16)$$
where the sum runs over all graphs $g$ with $2f'$ external fermions and
$l$ photons. $\omega (g)$ is the singular order of the graph $g$ which
is given by the simple expression [3]
$$\omega (g)=4-3f'-l.\eqno(17)$$
In (16) we have used the fact that the derivative increases $\omega$
by 1.

For $\omega <-1$, the inner sum in (16) contains no term, hence the
expression vanishes, which proves the desired divergence relation in
this case. There remains to investigate the possible cases of $\omega\ge -1$.
In virtue of (17), there are only the following four cases of this
kind:
$$\vbox{\halign{&$#$&$#$&$#$&$#$&$#$&$#$ \cr
              &(f',l)\>&=\>&(0,2)\>&(0,4)\>&(1,1)\>&(1,2)\cr
              &\omega&=&2&0&0&-1\cr
              &{\rm case}&:&{\rm I}&{\rm II}&{\rm III}&{\rm IV.}\cr}}
\eqno(18)$$
These cases must now be examined.
According to the lemma
in Sect.3.2 of ref.[2], we have only to consider connected diagrams.
However, combinations of field operators which differ from each other
in $d^\nu_{f'}$ may agree in $\d_\nu d^\nu_{f'}$, in virtue of the
identities (14) and (15).
\vskip 0.4cm
{\it Case I: $(f',l)=(0,2),\omega =2$ }
\vskip 0.4cm
This is vacuum polarization. Since $f'=0$, Eq.(16) does not contain any field
operator, so that we have to deal with numerical distributions only. We
write down the anomaly relation (16) for the $t$-distribution, assuming
the external photon operators at $x_1$ and $x_2$ for convenience:
$$\d_{1\nu}\Pi^{\nu\mu}(x_1,x_2;x_3,\ldots x_n)=\biggl(\sum_{ijk}K_{ijk}
\d_i^\alpha\d_{j\alpha}\d_k^\mu+\sum_kL_k\d_k^\mu\biggl)\delta^{n-1},
\eqno(19)$$
where $\delta^{n-1}=\delta(x_1-x_n)\cdot\ldots\delta(x_{n-1}-x_n)$.
Different from (16), $x_n$ is an inner vertex here.
The r.h.s. is the most general covariant local distribution with
$\omega=3$. We now claim:

{\bf Proposition 1.} The anomaly (19) can be restricted to the
following form
$$\d_{1\nu}\Pi^{\nu\mu}=\Biggl[K_1\sum_{i=3}^n \d_i^\alpha\d_{i\alpha}
\d_1^\mu+K_2\sum_{i=3}^n\d_i^\alpha\d_{1\alpha}\d_i^\mu$$
$$+K_3\d_1^\alpha\d_{2\alpha}\d_2^\mu+K_4\d_2^\alpha\d_{2\alpha}\d_1^\mu
+(K_3+K_4)\d_1^\alpha\d_{1\alpha}\d_1^\mu
+K_5\d_1^\alpha\d_{1\alpha}\d_2^\mu
+K_6\d_1^\alpha\d_{2\alpha}\d_1^\mu+K_7\d_1^\mu\Biggl]\delta^{n-1}.
\eqno(20)$$

{\it Proof}. Calculating the divergence of (19) with respect to $x_2$,
the result must be symmetric in $x_1, x_2$:
$$(\d_{2\mu}\d_{1\nu}\Pi^{\nu\mu})(x_1,x_2;\ldots)=(\d_{2\mu}\d_{1\nu}
\Pi^{\mu\nu})(x_2,x_1;\ldots).\eqno(21)$$
This implies $L_k=0$ for $k\ne 1$ and $K_{ijk}=0$ for $i,j,k>2$.
Furthermore, in virtue of (21), only the following $K_{ijk}$ can be $\ne
0$: (i) $K_{ij1}$, $K_{i1k}, K_{1jk}$, (ii) $K_{mpk}, K_{mjp},
K_{imp}$, (iii) $K_{lmp}$, where $l,m,p\le 2$, $i,j,k>2$.

We now use the property that (19) is symmetric in all inner vertices $x_3,
\ldots x_n$. This allows to express case (ii), depending
on one internal index $>2$ only, by derivatives with respect to
$x_1,x_2$, for example :
$$\sum_{k=3}^nK_{mnk}\d_k^\mu\delta^{n-1}=K_{mn}\sum_3^n\d_k^\mu
\delta^{n-1}=K_{mn}(-\d_1^\mu-\d_2^\mu)\delta^{n-1},\eqno(22)$$
because
$$\sum_{k=1}^n\d_k^\mu\delta(x_1-x_n)\cdot\ldots\delta(x_{n-1}-x_n)
=0.\eqno(23)$$
This reduces this case to case (iii). In case (i) the symmetry in the
inner vertices implies
$$K_{ij1}=K_{\pi i\pi j1},\quad K_{i1k}=K_{\pi i1\pi k}\eqno(24)$$
for all permutations $\pi$. This means that the diagonal elements $i=j$,
$i=k$ are equal and all off-diagonal elements also. By redefining these
constants, the latter sum over $i\ne j$ can be transformed into an
independent summation over $i,j=3,\ldots n$, which, by (23), is also
reduced to case (iii). There remains the summation over diagonal terms,
leading to the first two terms in (20). Finally,
in case (iii) we have $K_{111}=K_{122}+K_{221}$, $K_{222}=0$ due to (21).
Then we arrive at the remaining terms in (20).

In the anomaly (20) the derivative $\d_{1\nu}$ can now be taken out
$$\d_{1\nu}\Pi^{\nu\mu}=\d_{1\nu}\Biggl[g^{\mu\nu}K_1\sum_{i=3}^n
\d_i^\alpha\d_{i\alpha}+K_2\sum_{i=3}^n\d_i^\nu\d_i^\mu$$
$$+K_3(\d_1^\nu\d_1^\mu+\d_2^\nu\d_2^\mu)+ K_4g^{\mu\nu}(\w_1+\w_2)+
K_5\d_1^\nu\d_2^\mu+K_6\d_2^\nu\d_1^\mu+K_7g^{\mu\nu}\Biggl]
\delta^{n-1}.\eqno(25)$$
The square bracket is a polynomial of degree $\omega(\Pi^{\mu\nu})=2$
and has the
symmetry properties of $\Pi^{\mu\nu}$. Therefore it can be transformed
away by renormalization of
$\Pi^{\nu\mu}$ which completes the proof of case I.
\vskip 0.4cm
{\it Case II: $(f',l)=(0,4),\omega=0$}
\vskip 0.4cm
This is photon-photon scattering where we have again to deal with one
numerical distribution only. As in the proof of Prop.1 (21), the anomaly
can be expressed by derivatives with respect to the external coordinates
$x_1,\ldots x_4$, using the symmetry in the inner vertices:
$$\d_{1\nu}t^{\nu\mu\alpha\beta}(x_1,x_2,x_3,x_4;x_5,\ldots x_n)=$$
$$=\sum_{k=1}^4\Bigl(K_{k1}\d_k^\mu g^{\alpha\beta}+K_{k2}\d_k^
\alpha g^{\mu\beta}+K_{k3}\d_k^\beta g^{\mu\alpha}\Bigl)\delta^{n-1}
.\eqno(26)$$
Again, $x_n$ is an inner vertex here.
Since $\d_{1\nu}\d_{2\mu}\d_{3\alpha}\d_{4\beta}t^{\nu\mu\alpha\beta}$
must be symmetric in $x_1,\ldots x_4$, it follows that $K_{11}=K_{12}=
K_{13} =K$ and all other $K_{kj}=0$. Then
$$\d_{1\nu}t^{\nu\mu\alpha\beta}=K\d_{1\nu}\Bigl(g^{\nu\mu}g^{\alpha
\beta}+g^{\nu\alpha}g^{\mu\beta}+g^{\nu\beta}g^{\mu\alpha}\Bigl)
\delta^{n-1},\eqno(27)$$
and this anomaly can again be transformed away by renormalization of
$t^{\nu\mu\alpha\beta}$.

\vskip 0.4cm
{\it Case III: $(f',l)=(1,1),\omega =0$}
\vskip 0.4cm
We have one external photon operator in this case which is now
attached to $x_n$, which is the differentiation variable in agreement
with (16).
The decomposition of (16) leads to the following classes of diagrams:
(i) the vertex function with external
Fermi operators $\psq(x_i)\psi(x_j), 1\le i\ne j\le n-1$,
(ii) taking (14) and (15) into account, we have to include reducible
diagrams containing the self-energy $\Sigma$, (iii) there is an
additional class of reducible diagrams containing  the vacuum
polarization tensor $\Pi^{\mu\nu}$. The diagrams (ii) and (iii)
have external Fermi operators $\psq(x_i)\psi(x_i)$.
For fixed $x_i, x_j$, the anomaly relation for the $t$-distribution reads
$$\d_{n\nu}\Bigl[\,:\psq(x_i)\Lambda^\nu(\ldots)\psi(x_j):\, +\,:\psq(x_n)
\gamma^\nu S_F(x_n-x_i)\Sigma(\ldots)\psi(x_j):+$$
$$+\,:\psq(x_i)\Sigma(\ldots) S_F(x_j-x_n)\gamma^\nu\psi(x_n):\,+\,:\psq(x_i)
\Pi^{\nu\mu}(\ldots)D_F(x_j-x_i)\gamma_\mu\psi(x_i):\Bigl]=$$
$$=\,:\psq(x_i)\Bigl(
K_0{\bf 1}+K_i\ds_i+K_j\ds_j+K_n\ds_n\Bigl)\delta^{n-1}\psi(x_j)\,:.
\eqno(28)$$
Here we have used gauge invariance $\d_\nu\Pi^{\nu\mu}=0$ (case I) in
order $(n-1)$.

{\bf Proposition 2.} The anomaly (28) can be restricted to
$$K\,:\psq(x_i)\ds_n\delta^{n-1}\psi(x_j):.\eqno(29)$$

{\it Proof.} The total anomaly, i.e. the sum over all $1\le i\ne j\le
n-1$ of (28), is symmetric in $x_1,\ldots x_{n-1}$.
Instead of explicit symmetrization, it is simpler to consider (28) on
totally symmetric test functions $\fii(\xnn)$. Then the term with
$\ds_i\delta$ can be transformed by means of the Dirac equation as
follows
$$\int :\psq(x_i)\ds_i\delta^{n-1}\psi(x_j):\,\fii(\xnn)\,dx_1\ldots
dx_n=$$
$$=-\int :\psq(x_n)\gamma^\mu\psi(x_n):\,(\d_{i\mu}\fii)(x_n,\ldots
x_n)\,dx_n+\eqno(30)$$
$$+im\int :\psq(x_n)\psi(x_n):\,\fii(x_n,\ldots x_n)\,dx_n$$
$$=\int :\psq(x_i)\ds_k\delta^{n-1}\psi(x_j):\,\fii(\xnn)\,dx_1
\ldots dx_n+$$
$$+im\int :\psq(x_n)\psi(x_n):\,\fii(x_n,\ldots x_n)\,dx_n,$$
for $k\ne i,j,n$, and similarly for $\ds_j\delta$. That means
$\ds_i\delta$ can be substituted by $(\ds_1
+\ldots +\ds_{n-1})\delta/(n-1)$ plus the mass terms. The latter may be
included in $K_0$ in (28). Using (22), the anomaly assumes the form
$$:\psq(x_i)(K'_0{\bf 1}+K\ds_n)\delta^{n-1}\psi(x_j):.\eqno(31)$$

Now we use charge conjugation invariance
$$U_cT_n(\xnn)U_c^{-1}=T_n(\xnn).\eqno(32)$$
Since (28) is multiplied by the photon operator $A(x_n)$, which changes
sign under charge conjugation, the anomaly (28), (31) must also change
sign if the $C$-conjugated term of (28) is enclosed. This implies $K'_0=0$,
which completes the proof of the proposition.

The remaining anomaly (29) can be transformed away by renormalization of
the vertex function $\Lambda^\nu$.
\vskip 0.4cm
{\it Case IV: $(f',\,l)=(1,\,2),\omega=-1$}
\vskip 0.4cm
In this case we have two external Fermi and two photon operators. If the
former have coordinates $x_i, x_j$ and the latter $x_n$ and $x_{n-1}$,
the anomaly relation for the $t$-distribution is of the following form:
$$\d_{n\nu}\Bigl[\,:\psq(x_i)t_4^{\nu\mu}(\ldots)\psi(x_j):\,+\>\hbox{
\rm reducible terms}\Bigl]=$$
$$=K_{ij}\,:\psq(x_i)\gamma^\mu\psi(x_j):\,\delta^{n-1}.\eqno(33)$$
Here $t_4^{\nu\mu}$ is the contribution of the irreducible diagrams, the
reducible terms may have other arguments in the spinor operators, similar
to (28). Adding the $C$-conjugated equations,
the l.h.s. is even under charge conjugation, because it is multiplied by
two photon operators in the total $n$-point distribution $T_n$. The
anomalous terms on the r.h.s. add up to one local term with support
$x_i=\ldots =x_j=x_n$. The latter is odd under charge conjugation, hence,
the factor $K$ in front must vanish.
This completes the proof of gauge invariance.

It should be emphasized that gauge invariance in the sense of (3) is
weaker than the C-number Ward identities [2]. The latter are proven here
in case III between spinor fields only. To prove them in general needs
some input about the infrared behaviour, as the existence of the central
splitting solution in
the massive case [2]. We will return to this question elsewhere. But
condition (3) is all what is needed for the proof of unitarity [3]. The
method described here can also be used in Yang-Mills theories [7].
\vskip 0.8cm
{\bf References}\vskip 0.8cm
[1] J.S.Feldman, T.R.Hurd, L.Rosen, J.D.Wright,

QED: A Proof of Renormalizability,

Lecture Notes in Physics 312 (1988), Springer-Verlag

[2] M.D\"utsch, F.Krahe, G.Scharf, Nuovo Cimento 103 A, 903 (1990)

[3] G.Scharf, Finite Quantum Electrodynamics, Texts and  Monographs

in Physics, Springer-Verlag 1989

[4] H.Epstein, V.Glaser, Annales de l'Institut Poincar\'e A 19, 211
(1973)

[5] P.Blanchard, R.Seneor, Ann.Inst.H.Poincare 23, 147 (1975)

[6] T.R.Hurd, Comm.Math.Phys. 120, 469 (1989)

[7] M.D\"utsch, T.Hurth, F.Krahe, G.Scharf, Causal Construction of

Yang-Mills Theories I, Nuovo Cimento 106 A, 1029 (1993)

and II, Nuovo Cimento A (1994) to appear
\vskip 0.8cm

\vfill\eject\bye